# Water transport inside a single-walled carbon nanotube driven by temperature gradient

J. Shiomi and S. Maruyama

Department of Mechanical Engineering, The University of Tokyo
7-3-1 Hongo, Bunkyo-ku, Tokyo 113-8656, Japan

**Abstract.** In this work, by means of molecular dynamics simulations, we consider mass transport of a water cluster inside a single-walled carbon nanotube (SWNT) with the diameter of about 1.4 nm. The influence of the non-equilibrium thermal environment on the confined water cluster has been investigated by imposing a longitudinal temperature gradient to the SWNT. It is demonstrated that the water cluster is transported with the average acceleration proportional to the temperature gradient. Additional equilibrium simulations suggest that the temperature dependence of the potential energy of the confined water is sufficient to realize the transport. Particularly for the system with hydrophobic interface, the water-water intrinsic potential energy appears to play a dominant role. The transport simulations were also performed for a system with a junction between two different SWNTs. The results suggest that an angstrom difference in diameter may result in large barrier for water transported through a small diameter SWNT.

PACS: 61.46.Fg, 65.80.+n, 47.61.-k, 68.60.Dv

## 1. Introduction

Over the last decade, water confined in carbon nanotubes have attracted much attention as a realistic model of water confined in quasi-one-dimensional channels, a general case appearing in key systems in bioscience [1] and nanotechnology. Particularly, water inside a single-walled carbon nanotube (SWNT) serves as an ultimate realistic case of water under quasi-one-dimensional confinement. The confinement is expected to alter water transport properties [2, 3] and phase transition [4-7] from those of bulk water. Extensive studies have been also motivated by the growing expectation for micro-nanoscale manipulators and transporters, where characterization of molecular scale thermo-fluid dynamics under confinement is essential. These studies also serve to identify the dynamic interaction of water and carbon nanotubes, which is important in SWNT applications under aqueous environment.

The current report is designed to explore the possibility of transporting material along carbon nanotubes by using temperature gradient. The idea has been presented and demonstrated for solid materials namely for gold nanoparticle [8, 9] and outer layer of an double-walled carbon nanotube (DWNT) [10]. In the former reports, using molecular dynamics (MD) simulations, the mechanism has been discussed in connection with the interface interaction in analogy to the thermophoresis [8]. Furthermore, it is suggested that the phonons of the SWNT plays an important role to thrust the gold nanoparticle along the nanotube [9]. Barreiro et al [10] offers the first experimental observation of such phenomenon together with MD simulations, where the outer-tube of the DWNT, which is manufactured to be shorter than the inner-tube, is transported along the temperature gradient created by joule heating.

It is not clear if such phonon thrusting effect would be applicable for liquid cases since the pressure effects are expected to diffuse faster for liquid than solid. The aim here is to demonstrate the consequence in the case of liquid, namely water which is omnipresent in engineering environment. The realization of an efficient transport technique may motivate us to explore its use for nanoscale pumping devices. The demonstration is done by simulating a mass transport of a saturated water cluster locally confined in an SWNT with temperature gradient applied to the SWNT. The MD simulation incorporates realistic interfacial dynamics by modeling the carbon-carbon interaction dynamics with a potential function that has been shown to exhibit the phonon density of states of SWNTs with a sufficient accuracy [11-13]. Such model also allows us to investigate the influence of a junction between different SWNTs on the water transport, which is briefly demonstrated in this report.

## 2. Molecular dynamics simulations

Water molecules were expressed by the SPC/E potential [14]. The SPC/E potential is expressed as the superposition of Lennard-Jones function of oxygen-oxygen interaction and the electrostatic potential by charges on oxygen and hydrogen as follows,

$$\phi_{12} = 4\varepsilon_{OO}\left[\left(\frac{\sigma_{OO}}{R_{12}}\right)^{12} - \left(\frac{\sigma_{OO}}{R_{12}}\right)^{6}\right] + \sum_i\sum_j \frac{q_i q_j e^2}{4\pi\varepsilon_0 r_{ij}}, \qquad (1)$$

where $R_{12}$ represents the distance of oxygen atoms, and $\sigma_{OO}$ and $\varepsilon_{OO}$ are Lennard-Jones parameters. The Coulombic interaction is the sum of 9 pairs of point charges, where $r_{ij}$ denotes the distance between inter-molecular point charges. The potential function between water molecules and carbon atoms were represented by Lennard-Jones function of the distance between the oxygen in the water molecule and the carbon atom. The parameters for the Lennard-Jones potential were $\varepsilon_{CO}$=0.3136 kJ/mol and $\sigma_{CO}$=3.18 Å.

The carbon interactions were expressed by the Brenner potential function [15] in a simplified form [16], where the total potential energy of the system is modeled as,

$$E = \sum_i \sum_{j(i<j)} \left[V_R(r_{ij}) - B^*_{ij} V_A(r_{ij})\right]. \qquad (2)$$

Here, $V_R(r)$ and $V_A(r)$ are repulsive and attractive force terms, which take the Morse type form with a certain cut-off function. $B^*_{ij}$ represents the effect of the bonding condition of the atoms. As for the potential parameters, we employ the set that was shown to reproduce the force constant better (table 2 in Ref. 15). It has been shown that the formulation of potential function exhibits phonon dispersion with sufficient accuracy [11-13]. The inclusion of the lattice vibrations of carbon nanotubes enables us to incorporate the realistic heat transport from an SWNT to the water cluster. The thermal boundary conductance between the SWNT and the confined liquid water cluster was previously computed to be typically about 5 MW/m²K [17]. The cut-off distance of the Coulombic potential was set to be longer than the water cluster size. The velocity Verlet method was adopted to integrate the equation of motion with the time step of 0.5 fs.

As shown in figure 1, the simulated system consists of a water cluster of 192 molecules confined in a 64 nm long (10, 10) SWNT with a diameter of 1.38 nm. The length of the SWNT was determined to be long enough to achieve a steady transport of water inside the SWNT. After initially equilibrating the system at around 300 K, each end of the SWNT was cooled/heated by using Langevin thermostat to impose a temperature gradient along the nanotube with an average temperature of 300 K. As in the previous reports [11, 12, 18], the Langevin thermostat consists of a fixed layer and a phantom layer, which are both monolayer unit-cells. The phantom layer is placed between the fixed layer at the tube-end and the rest of the SWNT, and controlled by the Langevin equation. For the damping parameter of the Langevin equation, Debye temperature was chosen to be 2500 K, which is as high as that of diamond. The implementation of boundary conditions is an important issue which may influence the simulation results [19]. In the current study, it was confirmed that the time-averaged temperature profiles were always linear except for the temperature jumps between the thermostats and rest of the nanotube [17], which play minor role in the current study. Until the quasi-steady temperature gradient is achieved, the mean axial translational velocity of the water cluster was cancelled to maintain the center of mass at $z=0$. Once the quasi-steady thermal gradient is achieved with constant heat flux through the SWNT, the water cluster was released and the position and velocity of the center of mass was recorded. The temperature difference $T_h$-$T_c$ (figure 1) was varied to investigate the sensitivity of the transport velocity to the temperature gradient. Since the water cluster mass center is initially placed at $z=0$, the highest temperature the water cluster experiences is about 300 K.

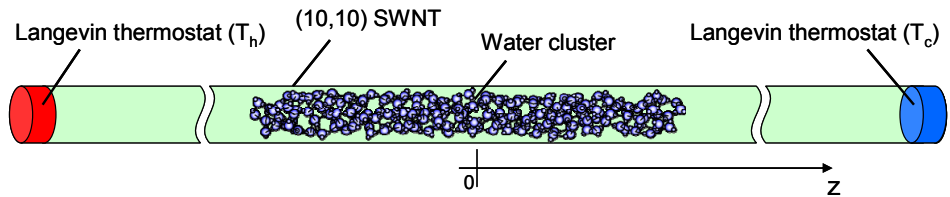

Figure 1: A schematic of the transport simulation. A cluster of 192 water molecules is placed inside a 64 nm long (10, 10) SWNT, and the temperature gradient is applied by heating and cooling the left and right ends, respectively, by Langevin thermostats.

**3. Water transport inside an SWNT by temperature gradient**

The results of the above molecular dynamics simulations clearly demonstrate the transport of water towards the cold end of the SWNT. The direction matches with the previous works on transport of sold materials [8-10]. Various phases of the transport are evident in the time history of the water cluster center of mass as shown in figure 2. After the initial phase, where the water cluster is pinned at $z=0$ (I), the water cluster accelerates at roughly constant acceleration (II) and, after a certain transient phase (III), eventually reaches a steady velocity (IV). For this case with a temperature gradient of 1.48 K/nm, the velocity reaches up to about 100 m/s. The water cluster eventually reaches the cold end and bounces back due to the increased potential outside the nanotube, but the thermal thrust keeps the cluster around the cold end (V). The time history of the mass center clearly reveals that the water cluster is driven with constant force which is balanced by the friction in the steady transport phase (IV). The friction is dominantly generated at the interface, judging from the plug-flow velocity distribution of the transported water.

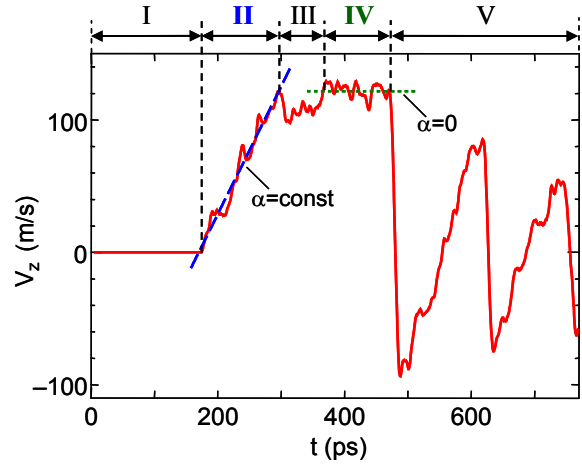

*Figure 2: The axial velocity time history $V_z(t)$ of the mass center of the water cluster. The imposed temperature gradient is 1.48 K/nm, and α denotes the corresponding acceleration. The time history is divided into, (I) initial phase with $V_z = 0$, (II) inertial phase with constant α, (III) transition phase, (IV) steady phase with constant α=0, and (V) the end phase with the cluster oscillating at the tube end.*

The transport simulations were performed for various temperature gradients ranging from 0.18 to 1.48 K/nm. The dependence of the driving force, calculated from the acceleration of the water cluster, on the temperature gradient is shown in figure 3. The acceleration is obtained by fitting the velocity time history in the acceleration phase (II) of the water transport (figure 2). Each point in the figure denotes an averaged value over 6 runs with different initial conditions. The error bars represent the random error of the 6 runs. It can be seen that the driving force increases proportionally with the temperature gradient within the extent of the error bars.

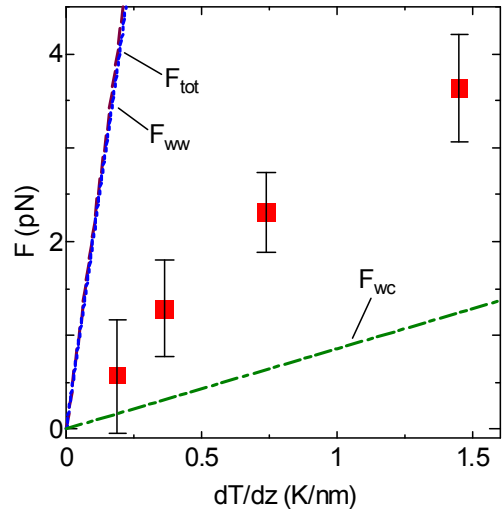

*Figure 3: Dependence of the driving force on the axial temperature gradient. Squares denote the results obtained from the inertial phase (Fig. 2) in the transport simulations. Each mark shows the average value of 6 simulations with different initial conditions. The accelerations $F_{ww}$, $F_{wc}$, and $F_{tot}$ calculated from the temperature dependence of the water-water, water-carbon, and total potential energy of the water cluster.*

Figure 3 suggested that the origin of the driving force lies in a quantity that has linear dependence on the temperature. Here, in course of considering the temperature dependence of the free energy, we take an energetic approach to gain insight into the mechanism of the transport. Since the free energy of the transported cluster depends on temperature, the state is expected to be led to more stable state, which is usually towards the cold side of the system. In a mind of enegetics, one could relate the temperature dependence of the potential energy and the force experienced by the water cluster. In case of water cluster inside a carbon nanotube, due to the hydrophobicity, it is expected that the water-water interaction plays dominant role compared with the water-carbon interaction, and therefore the intrinsic energetics may become more important than the interfacial ones.

Firstly, we identify the temperature dependence of the potential function of the water cluster, by performing isothermal MD simulations. In this case, the entire SWNT was temperature-controlled by the velocity scaling. Starting from 300 K, the SWNT was slowly cooled down to 230 K maintaining a quasi-equilibrium state. The average potential energies of water-water and water-carbon interactions were recorded along the simulations. As shown in figure 4, for both water-water and water-carbon interactions, the potential energy per molecule $\phi$ exhibit roughly linear dependences with gradients of $\partial \phi_{ww}/\partial T = 6.4 \times 10^{-2}$ kJmol$^{-1}$K$^{-1}$ and $\partial \phi_{wc}/\partial T = 2.7 \times 10^{-3}$ kJmol$^{-1}$K$^{-1}$, respectively. The temperature dependence of the water-water potential is dominantly determined by the Coulombic contribution ($8.5 \times 10^{-2}$ kJmol$^{-1}$K$^{-1}$) over the Lennard-Jones contribution ($-2.1 \times 10^{-2}$ kJmol$^{-1}$K$^{-1}$). Note that the hydrogen bonds network of the water in an SWNT is distorted by the confinement effect compared with that of bulk water [20]. It should be noted that for a (10, 10) SWNT, the lowest temperature in the current simulation (230 K) is higher than the freezing temperature of the ice-nanotube [6].

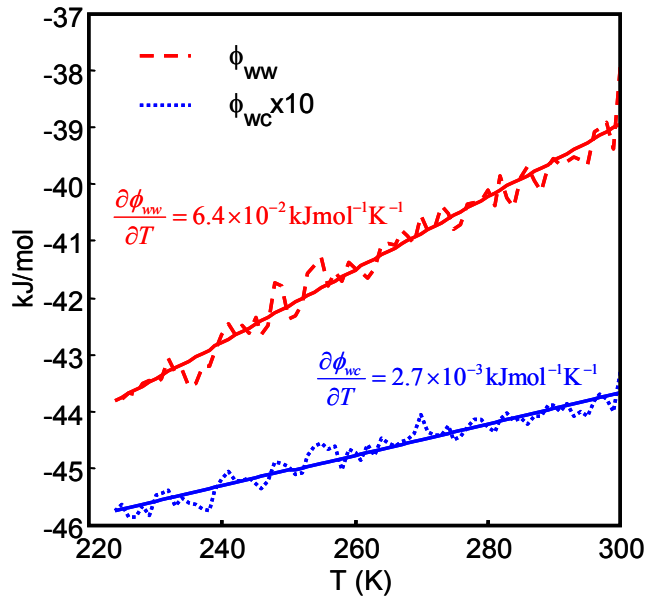

Figure 4: *The dependence of the potential energies on temperature. The data are plotted for water-water and water-carbon interaction potential energies.*

Now, we estimate the force that corresponds to the gradient of the potential energy with respect to the axial direction caused by the temperature gradient. Assuming that the water potential energy per unit length $\widetilde{\phi}$ can be expressed as a linear function of $z$, $\widetilde{\phi} = az + b$, the potential energy of the whole water cluster with length $L_z$ can be written as

$$\Phi = \int_{z_1}^{z_2} \widetilde{\phi} dz = aL_z z_G + bL_z, \quad (3)$$

where $z_G=(z_1+ z_2)/2$ is the mass center position of the water cluster.
Hence, the force contributing to the water cluster transport can be expressed as

$$F = \frac{\partial \Phi}{\partial z_G} = aL_z. \quad (4)$$

Since $a = \frac{\partial \widetilde{\phi}}{\partial z} = \frac{\partial \widetilde{\phi}}{\partial T}\frac{\partial T}{\partial z}$, we obtain

$$F = \frac{\partial \widetilde{\phi}}{\partial T}\frac{\partial T}{\partial z} L_z \quad (5)$$

By relating the expression to the isothermal MD simulation results using $L_z \partial \widetilde{\phi}/dT = N \partial \phi/dT$, with $N$=192 being the number of water molecules, temperature-gradient dependence of $F$ can be obtained as plotted in figure 3. The figure shows that the force obtained from the nonequilibrium transport simulations falls in between the two lines of $F_{ww}$ and $F_{wc}$. This suggests that the potential gradient of the water-water potential is enough to drive the cluster but not solely the water-carbon potential energy, due to the superior water-water potential energy to hydrophobicity water-SWNT interaction. While we use temperature gradient of the SWNT for the calculation, it would be more suitable to use the averaged water temperature since the temperature experienced by the water is different from the local SWNT temperature due to the finite interfacial thermal resistance. We have computed the temperature gradient experienced by the water cluster for the case with the largest temperature gradient (1.48 K/nm), where the discrepancy is expected to be largest, and obtained 0.68 K/nm i.e., the temperature gradient is overestimated by about a factor of two. However, it is small enough that the above discussion on the driving force remains the same.

**4. Influence of SWNT junction**

Finally, the influence of a junction of nanotubes with different diameters is briefly introduced. As shown in figure 5, a (9, 9) SWNT and a (8, 8) SWNT were smoothly connected using 5-membered and 7-membered rings at the junction [21]. In the same manner as the previous simulations for the smooth (10, 10) SWNT, the water transport simulations were performed by heating/cooling the end of the (9, 9)/(8, 8) SWNT. The temperature gradient here is 1.48 K/nm, which corresponds to the highest temperature gradient in the previous smooth SWNT simulations. In this case, the water cluster was initially placed off center at around z=-6 nm, inside the (9, 9) SWNT. After releasing the water cluster, the water cluster was transported towards the cold side i.e. towards with junction at a speed of approximately 90 m/s, similar to the value of the smooth (10, 10) SWNT case for the same temperature gradient. However, in this case, the cluster never crosses the junction, as shown in the trajectory and the sequent pictures (figure 6). Instead, when the water cluster reaches the junction, it oscillates due to the potential barrier at the junction counteracting the thrust from the temperature gradient. This is due to the fact that the water cluster is more stable inside the (9, 9) SWNT than inside the (8, 8) SWNT. Such difference has been already seen in the previous work [6], which showed that slight change in the diameter results in a significant variation of the potential energy and molecular structure. In the current case, the cluster does not flow into the (8, 8) SWNT, since the temperature gradient effect cannot overcome the potential energy barrier caused by the diameter difference, with the current temperature gradient. The current study does not discard the possibility of the contraction being a larger obstacle than the potential energy difference. More studies are underway to identify this aspect. Nevertheless, the current results suggest that a nanotube junction, often seen in synthesized SWNTs [22], may become a key obstacle. It should also be noted that such effect is expected to be particularly pronounce in the current small diameter SWNTs, and the influence is expected to become weaker as the diameter increases.

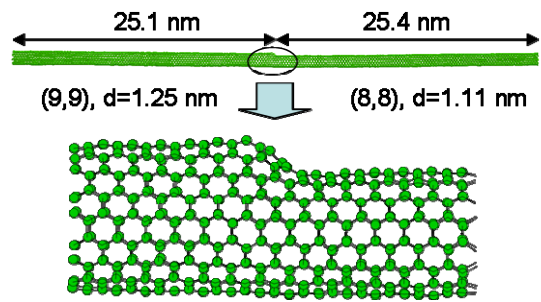

*Figure 5: A junction of (9, 9) and (8, 8) SWNTs.*

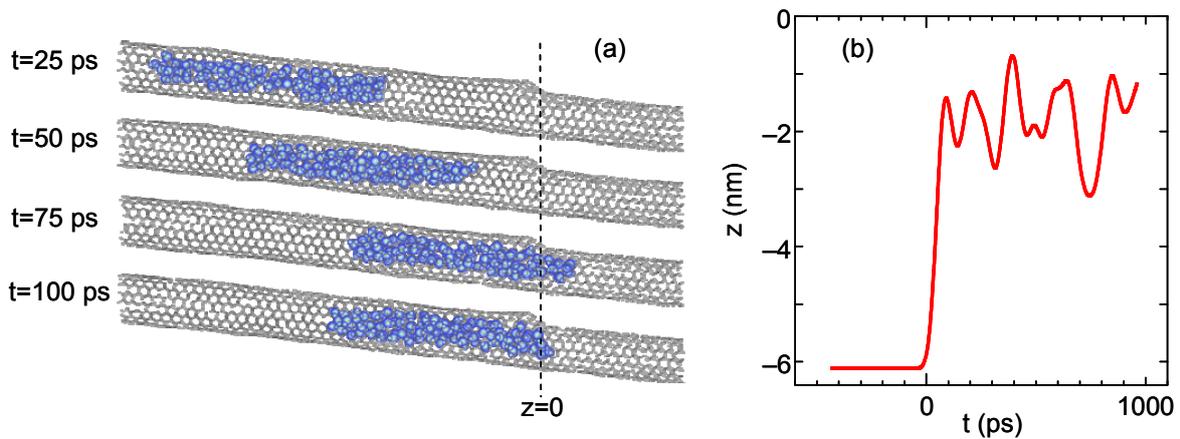

*Figure 6: The water cluster transport simulations with the (9, 9)-(8, 8) SWNT junction. A temperature gradient of 1.48 K/nm was applied to drive the water cluster in the (9, 9) SWNT towards the junction; (a) the time sequence images of the water cluster during the transport, and (b) the time history of the water cluster center of mass. The water cluster was initially (t<0) placed at z= -6 nm.*

## 5. Conclusions

We have demonstrated that the water cluster can be transported through a carbon nanotube by using temperature gradient. The energetic analysis suggests that the temperature dependence of the water-water interaction potential energy is sufficient to cause the thermally driven transport. By performing the transport simulations for SWNTs with a diameter varying junction, it was demonstrated that such a junction may become a detrimental obstacle on transporting water through carbon nanotube, especially for carbon nanotubes with small diameters.

**Acknowledgments**

This work is supported in part by Grants-in-Aid for Scientific Research 19051016 and 19860022.